\documentclass[journal]{journal}

\usepackage{cite}

\ifCLASSINFOpdf
\else
\fi

\usepackage{times}
\usepackage{graphicx}
\usepackage[bf,SL,BF]{subfigure}
\usepackage{pdfcomment}
\usepackage{booktabs}
\usepackage{hyperref}
\usepackage[algo2e,ruled]{algorithm2e}
\usepackage{caption}
\usepackage{color}
\usepackage[dvipsnames]{xcolor}
\usepackage{amsmath,amsthm,verbatim,amssymb,amsfonts,amscd,mathrsfs,mathtools}
\newtheorem{prop}{Proposition}
\usepackage{multirow}

\usepackage{hhline}
\usepackage{url}
\usepackage{physics}
\usepackage{bm}

\usepackage[none]{hyphenat}
\usepackage[justification=centering]{caption}
\usepackage[labelsep=space]{caption}
\usepackage[font=footnotesize]{caption}
\begin{document}

\title{Ordinal Regression with Fenton-Wilkinson Order Statistics: A Case Study of an Orienteering Race}

\author{Joonas~P\"a\"akk\"onen
\thanks{J. P\"a\"akk\"onen is with the Department of Informatics, School of Technology and Business Studies, Dalarna University, B\"orl\"ange, Sweden (e-mail: jpa@du.se).
}
}

\maketitle
\thispagestyle{empty}

\begin{abstract}
In sports, individuals and teams are typically interested in final rankings. Final results, such as times or distances, dictate these rankings, also known as \emph{places}. Places can be further associated with ordered random variables, commonly referred to as \emph{order statistics}. In this work, we introduce a simple, yet accurate order statistical ordinal regression function that predicts relay race places with changeover-times. We call this function the \emph{Fenton-Wilkinson Order Statistics} model. This model is built on the following educated assumption: individual leg-times follow log-normal distributions. Moreover, our key idea is to utilize Fenton-Wilkinson approximations of changeover-times alongside an estimator for the total number of teams as in the notorious German tank problem. This original place regression function is sigmoidal and thus correctly predicts the existence of a small number of elite teams that significantly outperform the rest of the teams. Our model also describes how place increases linearly with changeover-time at the inflection point of the log-normal distribution function. With real-world data from Jukola 2019, a massive orienteering relay race, the model is shown to be highly accurate even when the size of the training set is only 5\% of the whole data set. Numerical results also show that our model exhibits smaller place prediction root-mean-square-errors than linear regression, mord regression and Gaussian process regression.
\end{abstract}


\IEEEpeerreviewmaketitle

\section{Introduction}
\IEEEPARstart{C}{lassification} refers to machine learning methods where the target variable is a discrete class, while regression is typically associated with continuous variables. However, when the number of classes is large, yet discrete, it becomes challenging to make a distinction between classification and regression. \emph{Ordinal regression}, also known as \emph{ordinal classification}, refers to regression with a target that is discrete and ordered. It can thus be regarded as a hybrid mixture of both classification and regression.

Typical applications of ordinal classification include age estimation with an integer-valued target, advertising systems, recommender systems, and movie ratings. For additional insight into recent developments of related machine learning methods, the reader is kindly directed, \emph{e.g.}, to \cite{guti2016} for a survey on ordinal regression, and to \cite{raghu2020} for a survey on deep learning.

Ordinal regression lends itself especially well to ordered sets. In sports, all result lists are ordered sets with respect to results such as times, distances or points. Thus, for a given result, ordinal regression could predict the final rankings, \emph{i.e.}, the \emph{places} of teams or individual athletes. Here we conduct a case study of ordinal regression on the ranks of sorted sums of random variables of the duration of a relay. To be more specific, we study a large number of realizations of the changeover-times of an orienteering relay race.

We compare three widely-used regression schemes to an original ordinal classification method, the derivation of which is attributed to algebraic manipulations and well-known results concerning ordered random variables. Such random variables are known as \emph{order statistics}, and they represent a branch of mathematical statistics closely related to \emph{extreme value theory} (EVT). While sports analytics has seen several EVT applications for record values \cite{strand1998,spearing2019}, in this work we do not focus on extreme values but rather order statistics in general.

As an underpinning educated assumption, we say that individual leg-times are log-normal. We furthermore assume the log-normality of changeover-times, which is due to the Fenton-Wilkinson approximation \cite{fenton1960,wilk1967,cobb2012}. We also note that while there exist explicit expressions for the expectations of log-normal order statistics \cite{nadarajah2008}, for our purposes these expressions are unnecessary as scaling the log-normal distribution function directly produces a place predictor.

According to the principle of maximum entropy \cite{jaynes1957}, one could argue that the amount of uncertainty in the relay system increases with time, and that changeover-times thus tend to follow a maximum entropy distribution, such as the log-normal distribution. In practice, though, the log-normality assumption follows from the observation that marathon finish-times exhibit log-normality \cite{allen2014}. We show that a log-normal shape also fits orienteering data, which is to be expected given that both marathon running and orienteering are endurance sports.

Unlike prediction models for individual marathon race finish-times \cite{ruizmayo2016,esteve2019}, here we consider orienteering relay team place prediction. As a distinctive element of our work, rather than predicting times, we are interested in predicting places. It is often the place that is the hard, quantitative result that many teams wish to minimize. Thus, place prediction is of particular interest.

The main contributions of this work are the introduction and the validation of what we refer to as the Fenton-Wilkinson Order Statistics (FWOS) model. For a case study of an orienteering race with real-world data, numerical results show that FWOS accurately predicts places even with very few training examples. Further, FWOS plots correctly illustrate that place increases sigmoidally with changeover-time.

\section{System Model}
Consider an orienteering relay race. Let $n$ denote the number of finishing teams as we ignore disqualified and retired teams. There are $m$ runners on each team and each runner runs one \emph{leg}. Each leg is immediately followed by another at a \emph{changeover} until the end of the relay.

\emph{Leg-time} is the time result of an individual runner. Leg-times correspond to independent but not identically distributed random variables $Z_i$ with $i\in\mathbb{N}_m\coloneqq\{1,2,\dots,m\}$.

\emph{Changeover-time} is a team's cumulative time after $l$ legs. Changeover-time random variable $T^{(l)}$ after leg $l\in\mathbb{N}_m$ is defined as the sum of the first $l$ leg-times, \emph{i.e.},
\begin{align}\label{eq:lognsum}
T^{(l)} \coloneqq \sum_{i=1}^l Z_i.
\end{align}
Note that, technically, there is no changeover $l=m$ since ``changeover" $m$ is the finish. Also note that the final team finish-time result list of the relay race is a length-$n$ sample of $T^{(m)}$ sorted in ascending order.

In this article, \emph{place} always refers to the final team ranking after all the $m$ legs. The changeover-time of the team that arrives at changeover $l$ as the $r^\text{th}$ team out of $n$ teams is the realization of random variable $T_{r:n}^{(l)}$. Such a sorted random variable is called an \emph{order statistic}.

Changeover-time order statistic $T_{r:n}^{(l)}$, with changeover ranking $r\in\mathbb{N}_n\coloneqq\{1,2,\dots,n\}$, satisfies
\begin{align*}
T_{1:n}^{(l)} \leq T_{2:n}^{(l)} \leq \dots \leq T_{r:n}^{(l)} \leq \dots \leq T_{n:n}^{(l)}.
\end{align*}
Finish-time order statistic $T_{r:n}^{(m)}$, with place $r\in\mathbb{N}_n$, satisfies
\begin{align*}
T_{1:n}^{(m)} \leq T_{2:n}^{(m)} \leq \dots \leq T_{r:n}^{(m)} \leq \dots \leq T_{n:n}^{(m)}.
\end{align*}
Especially note that $T_{1:n}^{(m)}$ is the total time of the winning team.

Let $c<n$ denote the number of training observations. For each leg $l$, the training observations are chosen uniformly at random from the realizations of all $n$ changeover-times and their corresponding places. Hence, we are given $c$ realizations of $T^{(l)}$, \emph{i.e.}, a changeover-time training vector $$\bm{t}_l=\left(t_1^{(l)},\dots,t_c^{(l)}\right)\in \mathbb{R}_+^c$$ at changeover $l$, and the corresponding place training vector
\begin{align*}
\bm{r}=\left(r_1,\dots,r_c\right)\in B_c,
\end{align*}
where $B_c\coloneqq\{S\subset\mathbb{N}_n:|S|=c\}$ denotes the set of all proper $c$-subsets of $\mathbb{N}_n=\{1,2,\dots,n\}$. Hence, $\bm{t}_l$ are realizations of $T^{(l)}$ and $\bm{r}$ are the corresponding place examples.

\textbf{Problem formulation:} Let $t_i^{(l)}$ denote a changeover-time with index $i$ at changeover $l$ and $r_i$ denote the corresponding place. Our primary task is to find a place predictor function $\Upsilon^{(l)}:\mathbb{R}_+\to \mathbb{N};t\mapsto\Upsilon^{(l)}(t)$ that satisfies
\begin{align}\label{eq:yhx}
\Upsilon^{(l)}{\left(t_i^{(l)}\right)}\approx r_i
\end{align}
as accurately as possible. We want the prediction error, \emph{i.e.}, the approximation error of \eqref{eq:yhx}, to be as small as possible. In this work, we use the RMSE loss function to measure this prediction error, as will be discussed later in Section \ref{sec:rmse}.

\section{Regression Models}\label{sec:regs}
\subsubsection{Linear Regression} This regression model refers to the traditional Ordinary Least Squares (OLS) regression rounded to the nearest integer. OLS finds the intercept and slope that minimize the residual sum of squares between the observed targets and the targets predicted by the linear approximation. Fitting a straight line to the data reflects an initial, uneducated guess that place increases linearly with changeover-time.

\subsubsection{Gaussian Process (GP) Regression} A GP is a nonparametric model that can manage exact regression up to a million data points on commodity hardware \cite{wang2019}. For a pair of training vectors $(\bm{t}_l,\bm{r})$, a GP is defined by its \emph{kernel} function $k(\cdot,\cdot)$, a $c{\times}c$ kernel matrix $\bm{K}_{\bm{t}_l \! \bm{t}_l}$ with covariance values for all training pairs, and a $c$-dimensional vector $\bm{k}_{\bm{t}_l t}$ with evaluations of the kernel function at training point vector $\bm{t}_l$ and $t$. A Gaussian process predicts an arbitrary, unknown function $g(\cdot)$.

For kernel matrix $\widehat{\bm{K}}_{\bm{t}_l \! \bm{t}_l} = \bm{K}_{\bm{t}_l \! \bm{t}_l} + \sigma_0^2 I$, with additive Gaussian noise with zero mean and variance $\sigma_0^2$, the expected value of the zero mean GP predictive posterior distribution with a Gaussian likelihood is $\mathbb{E}{\left(g(t) \! \mid \! \bm{t}_l, \bm{r}\right)} = \bm{k}_{\bm{t}_l t}^\intercal\widehat{\bm{K}}_{\bm{t}_l \! \bm{t}_l}^{-1}\bm{r}$ \cite{rasmussen2006}. Hence, we define the GP place predictor as
\begin{align*}
\Upsilon_{\text{GP}}^{(l)}(t) \coloneqq \big[\bm{k}_{\bm{t}_l t}^\intercal\widehat{\bm{K}}_{\bm{t}_l \! \bm{t}_l}^{-1}\bm{r}],
\end{align*}
where $[\cdot]$ denotes rounding to the nearest integer.

For numerical implementations of exact GP, we utilize the readily available GPyTorch Python library with a radial basis function (RBF) kernel exactly as in the ``GPyTorch Regression Tutorial" in \cite{gpytorch} as a black box solution.

\subsubsection{Mord Regression} This regression model refers here to the regression-based model from the readily available Python mord package for ordered ordinal ridge regression. It overwrites the ridge regression function from the scikit-learn library and uses the (minus) absolute error as its score function \cite{mord, fabian2015}. For numerical implementation, we use the mord package exactly as in \cite{mord} as a black box solution.

\subsubsection{Fenton-Wilkinson Order Statistics (FWOS) Regression} Let $[\cdot]$ denote rounding to the nearest integer, let $\Phi(\cdot)$ denote the cumulative distribution function (c.d.f.) of the standard normal distribution, let $\log(\cdot)$ denote the logarithm, let $(\hat{\mu}_l,\hat{\sigma}_l)$ be the maximum likelihood estimates (MLE) of the log-normal parameters, and let $r_{(c)}$ denote the maximum of $c$ place observations $\{r_i\}$ with $i\in\mathbb{N}_c\coloneqq\{1,2,\dots,c\}$.

The FWOS regression model is defined as follows.

\begin{prop}
For $c$ pairs  of random changeover-time--place training observations, the FWOS regression function
\begin{align}\label{prop:prop}
\Upsilon_{\text{FWOS}}^{(l)}(t) \coloneqq \Bigg[\Phi\left(\frac{\log t - \hat{\mu}_l}{\hat{\sigma}_l}\right)\left(1+\frac1c\right)r_{(c)}\Bigg]
\end{align}
predicts place with changeover-time $t=t_l$ at changeover $l$.
\end{prop}

The derivation of \eqref{prop:prop} is deferred to the Appendix.

Loosely speaking, Proposition 1 states that FWOS approximates place with the expected changeover-ranking for a given changeover-time. We anticipate that this approximation holds to a satisfactory degree and that it improves with $l$.

\section{Numerical Results}\label{sec:numres}
Real-world data are acquired from the publicly available results of the prestigious annual orienteering relay Jukola, where there are $m=7$ runners on each team. We specifically use the results of Jukola 2019 \cite{jukola2019}, where there are $n=1653$ teams. The regression models are trained for two cases: 1) for $c=1322$ $\left(c/n\approx80\%\right)$, and 2) for $c=82$ $\left(c/n\approx5\%\right)$, randomly chosen pairs of changeover-time--place training observations for each of the seven legs. In both cases, the rest of the data are used for testing the regression models.

\begin{figure*}[h!]
  \centering
  \includegraphics[width=.81\linewidth]{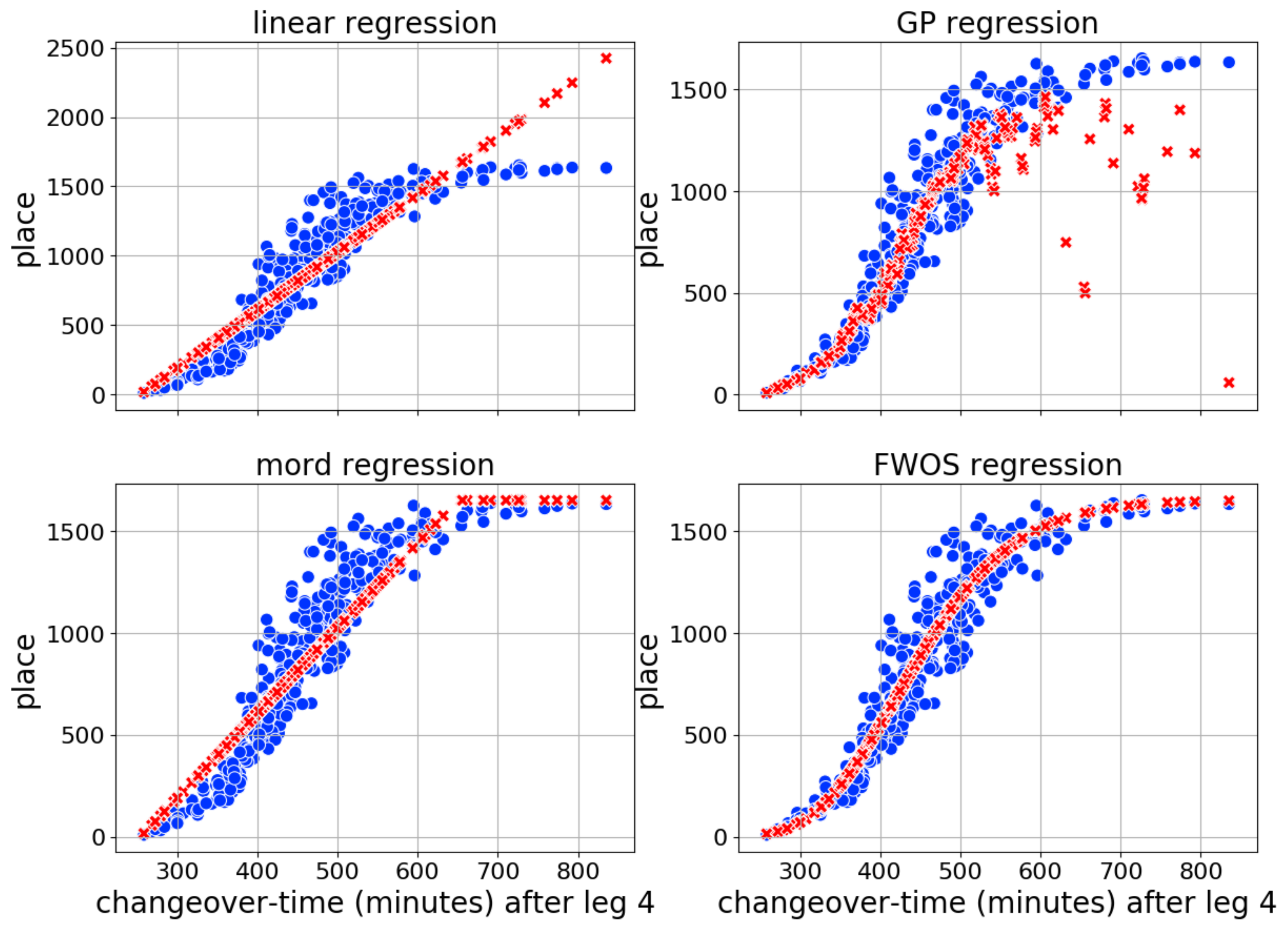}
  \caption{Place prediction (red crosses) against changeover-time at changeover $l=4$ with training set size 80\% and test set (blue circles) size 20\% for all the four tested regression models. FWOS regression captures the sigmoidal nature of the data.}
\label{fig:leg4ts20}
\end{figure*}

\begin{figure*}[h!]
  \centering
  \includegraphics[width=.81\linewidth]{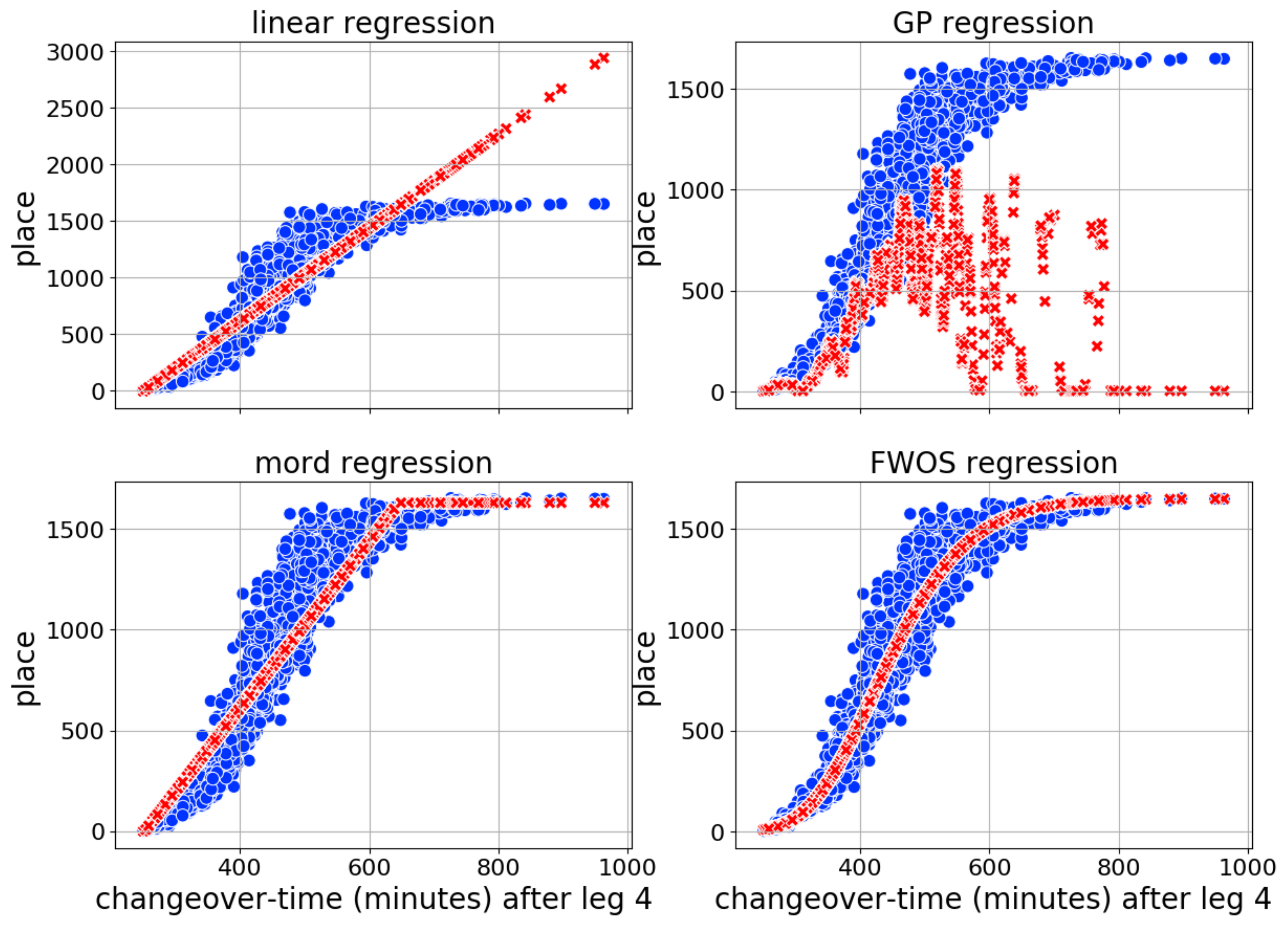}
  \caption{Place prediction (red crosses) against changeover-time at changeover $l=4$ with training set size 5\% and test set (blue circles) size 95\% for all the four tested regression models. The prediction performance of FWOS is similar to that of Fig. \ref{fig:leg4ts20} despite the significantly smaller number of training points.}
  \label{fig:leg4ts95}
\end{figure*}

\subsection{Curve Fits}
Fig. \ref{fig:leg4ts20} plots place against changeover-time (in minutes) after leg $l=4$ for the predictions (red crosses) and the test set (true values, blue circles) for all the four regression models for a random training set, the size of which is $80\%$ of the data set.

Linear regression and GP regression provide reasonable predictions for a large portion of the test set, but behave poorly when time is large. For average-performing teams, with approximately $350$ to $550$ minutes of elapsed time at the $4^{\text{th}}$ changeover, place seems to grow linearly with time.

Mord regression captures the effect where place saturates for large values of time, but fails to provide a smooth transition. FWOS regression, unlike the other models, indeed exhibits the smooth sigmoidal behavior of the data.

Interestingly, the red FWOS curve in Fig. \ref{fig:leg4ts20} suggests that \emph{there exists a rather small number of elite teams that ``pull away'' from the rest of the teams}, as suggested by the convex part of the curve, while extremely slow teams fall far behind the rest, as suggested by the concave part of the curve. If the curve was convex everywhere, no teams would fall far behind, while if the curve was concave everywhere, there would be no elite teams that distinctively pull away from the rest.

In Fig. \ref{fig:leg4ts95}, there are significantly fewer training data compared to Fig. \ref{fig:leg4ts20}, namely, $5\%$ compared to $80\%$. Yet, linear, mord and FWOS regression provide similar fits compared to those of Fig. \ref{fig:leg4ts20}, whereas GP greatly suffers from the lack of training data especially with high values of changeover-time.

The poor performance of GP regression may be due to, \emph{e.g.}, suboptimal hyperparameter values. While it is true that optimizing the GP method could improve its performance, it is extremely unlikely that an optimal GP could significantly outperform FWOS and we thus leave optimizing the GP method outside the scope of this work. Similar reasoning renders optimizing the mord method unnecessary.

\subsection{Root-Mean-Square Errors (RMSEs)}\label{sec:rmse}
To measure the error between place prediction $\Upsilon^{(l)}(t_i)$ and the corresponding true value $r_i$ of a test set with data point indices $i\in\mathbb{N}_v$, where $v = n - c$ is the size of the test set, we use the root-mean-square error (RMSE)
\begin{align}\label{eq:rmse}
\text{RMSE}{\left(\Upsilon^{(l)}\right)}\coloneqq\sqrt{\frac{1}{v}\sum_{i=1}^{v} \big(\Upsilon^{(l)}(t_i)-r_i\big)^2}
\end{align}
as a functional. We anticipate that \eqref{eq:rmse} decreases with $l$ because the chances of overtaking diminish as the relay progresses\footnote{For the same reason, the relay can be regarded as a composition of random permutations $\pi_l\in\text{Perm}(\mathbb{N}_n)$ of an ordered place set $\mathbb{N}_n$, where permutation $\pi_l$ at changeover $l$ ``approaches" the identity permutation as $l$ increases.}.

\begin{figure}[h!]
  \centering
  \includegraphics[width=.95\linewidth]{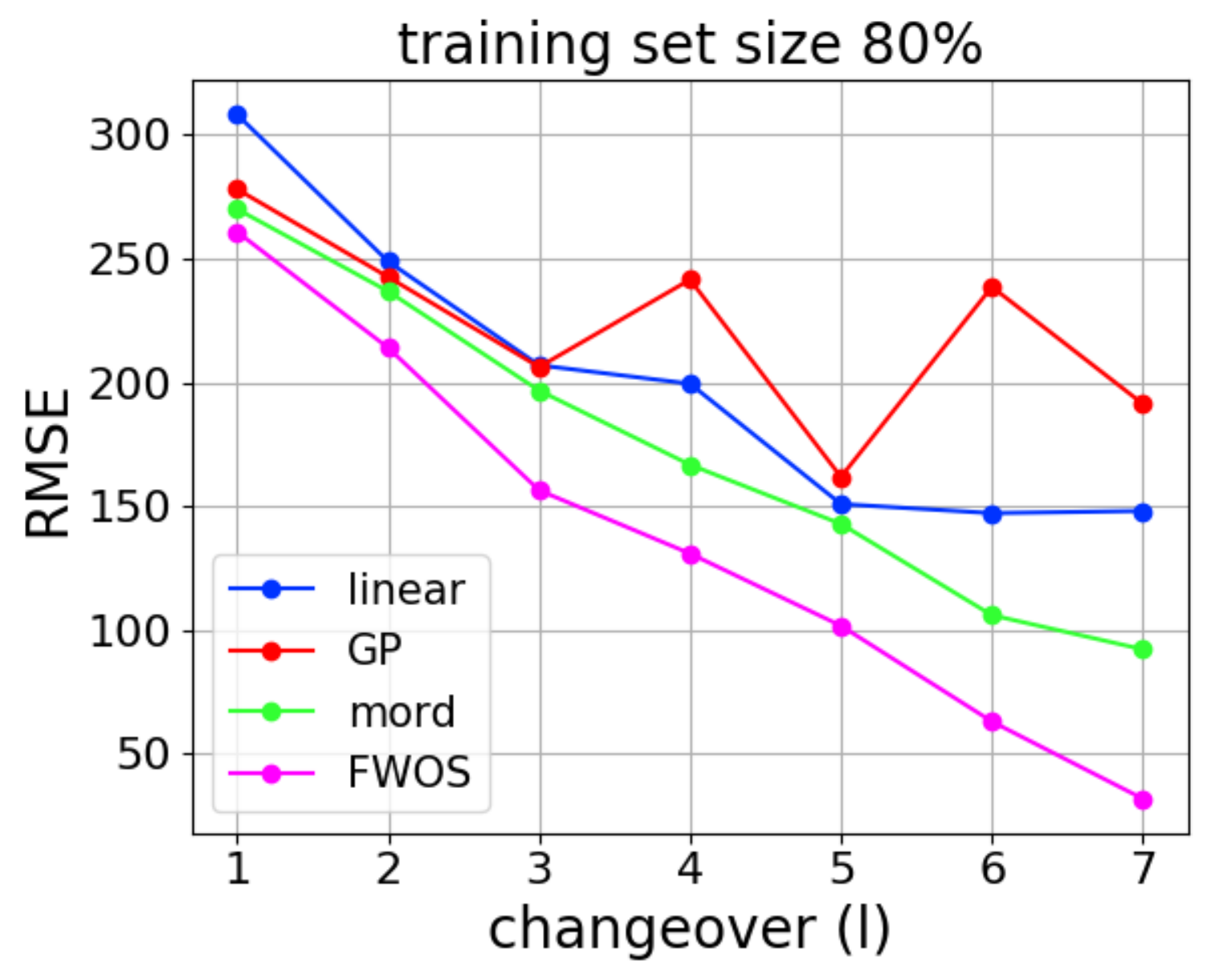}
  \caption{RMSEs for each changeover ($l$) with training set size $80\%$. FWOS displays the lowest RMSE values, and the FWOS RMSE curve seems to decrease linearly with changeover $l$.}
\label{fig:rmse20}
\end{figure}

\begin{figure}[h!]
  \centering
  \includegraphics[width=.95\linewidth]{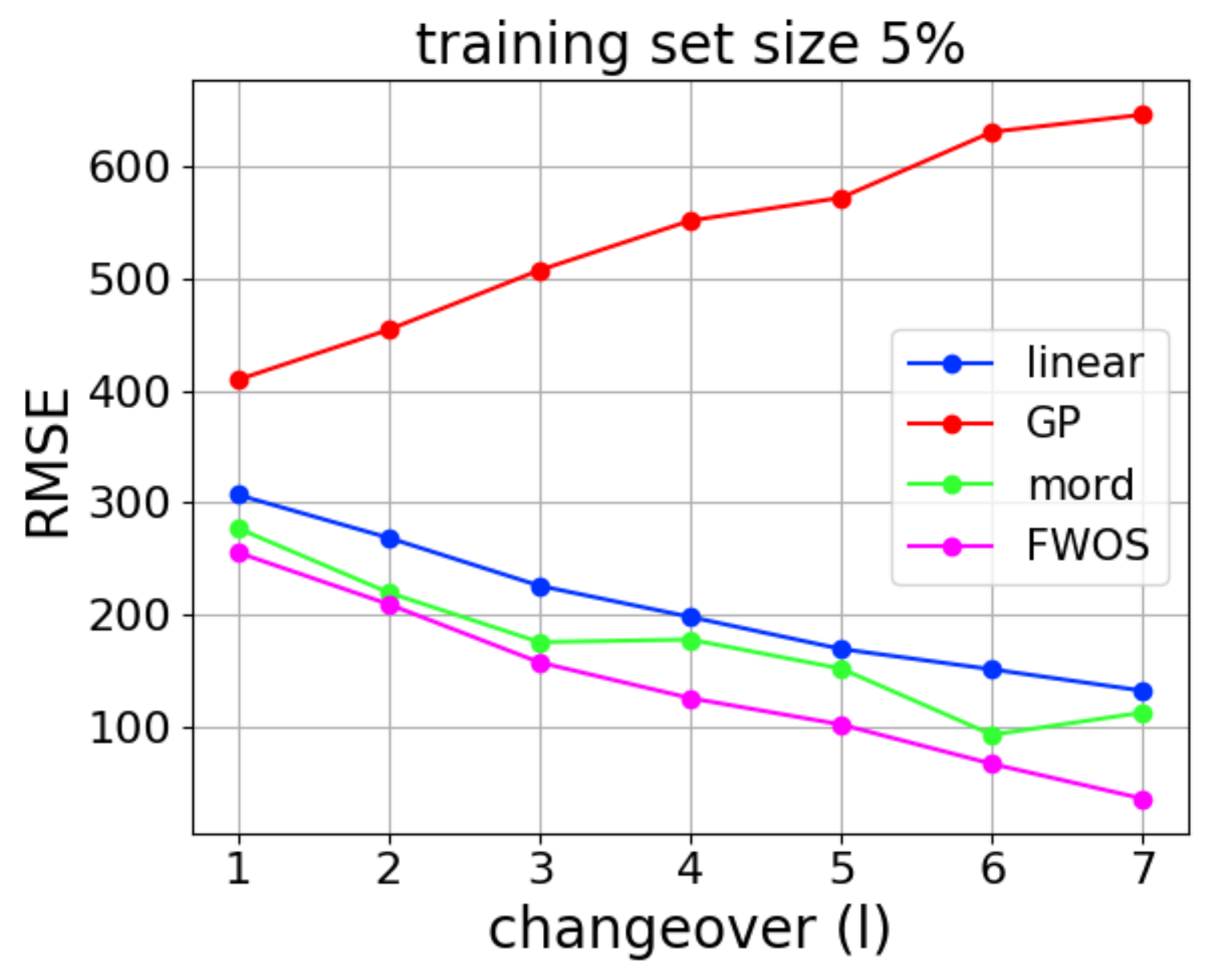}
  \caption{RMSEs for each changeover ($l$) with training set size $5\%$. FWOS regression provides the best results for all changeovers. The error performance of FWOS is comparable to that of Fig. \ref{fig:rmse20} despite the small amount of training data.}
\label{fig:rmse95}
\end{figure}

Fig. \ref{fig:rmse20} plots the RMSEs after each of the $m=7$ changeovers for a small random test set of $v=331$ test points (a large training set with $c/n\approx0.80$ as in Fig. \ref{fig:leg4ts20}), while Fig. \ref{fig:rmse95} plots the RMSEs for a large random test set of $v=1571$ test points (a small training set with $c/n\approx0.05$ as in Fig. \ref{fig:leg4ts95}). We notice similar RMSEs for both training set sizes for linear, mord and FWOS regression. However, when the training set is small, the error performance of the GP regression model unexpectedly deteriorates with $l$, as shown in Fig. \ref{fig:rmse95}.

Figs. \ref{fig:rmse20} and \ref{fig:rmse95} show that FWOS exhibits the lowest RMSE values when compared to the other models. However, the RMSEs are not zero even for the $7^{\text{th}}$ changeover, \emph{i.e.}, after the \emph{anchor leg} when the team finishes. This is due to the imperfections of the regression models and the random fluctuations of the data.

\subsection{Log-normal Statistics}
To study the full data set, Table \ref{table:modes} tabulates leg-distances, changeover-distances and log-normal statistics including the changeover-time mean $w=\mathbb{E}{\left[T^{(l)}\right]}=\exp(\mu_l+\sigma_l^2/2)$ and the changeover-time mode $u=\exp(\mu_l-\sigma_l^2)$ for changeover $l$. The changeover-specific log-normal parameters $\mu_l$ and $\sigma_l$ are found through maximum likelihood estimation (MLE) over the whole data set.

\begin{table}[h]
    \centering
\captionsetup{justification=centering}
\caption{\footnotesize \normalfont\scshape \\Log-Normal Statistics of Jukola 2019 Changeover-Times}
\begin{tabular}{@{}crrcccc@{}} \toprule
$l$ & $s$ & $\sum s$ & $w$ & $\Delta w$ & $u$ & $\Delta u$ \\ \midrule
        1 & 10.7 & 10.7 & 107.5 & 107.5 & 99.5 & 99.5  \\
        2 & 10.4 & 21.1 & 219.4 & 111.9 & 203.5 & 104.4  \\
        3 & 13.1 & 34.2 & 355.5 & 136.1 & 331.7 & 128.2 \\
        4 & 7.2 & 41.4 & 452.1 & 96.6 & 419.6 & 87.9 \\
        5 & 7.7 & 49.1 & 555.5 & 103.4 & 513.3 & 93.7 \\
        6 & 11.0 & 60.1 & 682.8 & 127.3 & 633.4 & 120.1 \\
        7 & 12.8 & 72.9 & 815.4 & 132.6 & 760.7 & 127.3 \\ \bottomrule
\end{tabular}
\label{table:modes}
\end{table}
Further, in Table \ref{table:modes}, $s$ denotes the leg-distance (in kilometers) and $\sum s$ denotes the cumulative changeover-distance (in kilometers) covered in total at changeover $l$. Also the change in the mean $\left(\Delta w\right)$ and the change in the mode $\left(\Delta u\right)$ compared to the previous changeover are shown.

Here the mode is a \emph{rising point of inflection}, \emph{i.e.}, a point where the $2^{\text{nd}}$ derivative of the c.d.f. fit changes its sign from positive to negative. Interestingly, and loosely speaking, at this point place increases linearly with changeover-time. Figs. \ref{fig:leg4ts20} and \ref{fig:leg4ts95} suggest that the FWOS mode for changeover 4 is around $420$ (minutes). Around that time teams arrive at the changeover at approximately constant intervals.

Table \ref{table:modes} reveals that changeover 3 yields the largest increases in the changes of both changeover-time means and changeover-time modes. Therefore, we may argue that leg 3 (colloquially: \emph{the long night}) is the most important leg.

The $3^{\text{rd}}$ leg in Jukola 2019 was only slightly longer than the anchor leg, but for the faster teams the $3^{\text{rd}}$ leg is a night leg as opposed to the anchor leg which is run in daylight. Night orienteering is typically somewhat slower than orienteering in daylight. However, for the slower teams, the $3^{\text{rd}}$ leg is in practice a dawn leg, or even a day leg, rather than a night leg.

\subsection{Additional Remarks}
The leg-distance of a certain leg is not exactly equal for each team as all the legs in Jukola are \emph{forked}: not all teams visit exactly the same control points on any given leg. In this manner, individual orienteering is enforced even in packs of runners. However, running in packs is common especially on the $1^{\text{st}}$ leg and after restarts. Pack running may distort leg-time distributions and thus erode our prediction performance. Orienteers typically arrive at changeovers in bursty clusters -- a phenomenon not captured by the smooth FWOS model.

\section{Conclusions}
This work has shed light on the numerical nature of relay races. We have introduced the Fenton-Wilkinson Order Statistics (FWOS) model in order to predict discrete places with continuous changeover-times. A real-world case study of an orienteering relay race has verified the accuracy of FWOS even with few training data. Based on these results, we advocate properly scaled log-normal c.d.f. fits for both \emph{place against leg-time} plots and \emph{place against changeover-time} plots. We also conjecture that, \emph{e.g.}, ultramarathons exhibit log-normal characteristics. Our results may further bring better understanding of pacing and pack clustering in large-scale endurance running sporting events.

\appendix[Derivation of FWOS Regression Function]
Let us make the following two well-educated assumptions.

\textbf{Assumption 1:} Each individual leg-time $Z_i$ is log-normal.

\textbf{Assumption 2:} Each changeover-time $T^{(l)}$ is log-normal.

The log-normal distribution often appears in sciences \cite{limpert2001}. Assumption 1 is based on the log-normality of travel time, such as vehicle travel time \cite{chen2018}, and, more importantly with regard to an endurance running application, marathon finish-times exhibit a log-normal shape \cite{allen2014}.

Assumption 2 concerns the sum of log-normal random variables as described in \eqref{eq:lognsum}. While it is widely-known that a closed-form expression does not exist for the density of a log-normal sum, it is commonly approximated by the partly folkloric \emph{Fenton-Wilkinson (FW) method} \cite{fenton1960,wilk1967,cobb2012}. This method models log-normal sums with another log-normal random variable. $Z_i$ and $T^{(l)}$ are thus both assumed to be log-normal but not identically distributed as their log-normal parameters are different except for the special case when $i=l=1$.

We can now derive the FWOS regression predictor function $\Upsilon_{\text{FWOS}}^{(l)}(\cdot)$ defined in \eqref{prop:prop}. To achieve this, we utilize the following two well-known preliminary tools in elementary probability theory.

\textbf{Tool 1:} Let $W$ denote a random variable that follows the standard uniform distribution $U(0,1)$ and let $T^{(l)}$ follow distribution $F$. Let $W_{r:n}$ denote the $r^{\text{th}}$ order statistic of a length-$n$ sample of $W$. The $r^{\text{th}}$ order statistic of a length-$n$ sample of $T^{(l)}$ has the same distribution as the inverse cumulative distribution function (c.d.f.) of $F$ at $W_{r:n}$

\textbf{Tool 2:} The $r^{\text{th}}$ standard uniform order statistic follows Beta($r,n - r + 1$). Therefore, the expected value of $W_{r:n}$ is $\mathbb{E}(W_{r:n}) = r/(n+1)$.

The inverse c.d.f. of $F$ is known as the \emph{quantile function} $Q_{F}(\cdot)$. Tool 1 can be thus expressed as
\begin{align}\label{eq:tool1}
T_{r:n}^{(l)} \stackrel{\text{d}}{=} Q_{F}(W_{r:n}),
\end{align}
where ``$\stackrel{\text{d}}{=}$" reads ``has the same distribution as". Hence, applying Tool 2 to \eqref{eq:tool1} yields the expected value\footnote{Explicit expressions for the expected values of $T_{r:n}^{(l)}$, \emph{i.e.}, the expected changeover-times for changeover-ranking $r$ out of $n$ teams at changeover $l$, can be found through \cite[Theorem~1]{nadarajah2008}. However, finding such expected values is unnecessary for our specific purposes.} of $T_{r:n}^{(l)}$ as
\begin{align}\label{eq:tool2}
\mathbb{E}(T_{r:n}^{(l)}) = Q_{F}{\left(\frac{r}{n+1}\right)}.
\end{align}
Let $F$ be the log-normal distribution with c.d.f. $F_{T^{(l)}}(\cdot)$. Now \eqref{eq:tool2} directly implies $F_{T^{(l)}}{\left(\mathbb{E}(T_{r:n}^{(l)})\right)} = r/(n + 1)$ and hence
\begin{align}\label{eq:rexact}
F_{T^{(l)}}{\left(\mathbb{E}(T_{r:n}^{(l)})\right)}(n + 1) = r.
\end{align}
For large $n$, as in our case study, it is fair to assume that $\forall t\in\mathbb{R}_+, \exists r\in\mathbb{N}_n$ such that
\begin{align}\label{eq:approxt}
\mathbb{E}{\left(T_{r:n}^{(l)}\right)}\approx t.
\end{align}
Combining \eqref{eq:rexact} and \eqref{eq:approxt}, we arrive at $F_{T^{(l)}}{\left(t\right)}(n + 1) \approx r$, which resembles \eqref{eq:yhx} as desired.

The log-normal c.d.f. is
\begin{align}\label{eq:logncdf}
F_{T^{(l)}}{(t)} = \Phi\left(\frac{\log t - \mu_l}{\sigma_l}\right),
\end{align}
where
\begin{align*}
\Phi(x)=\frac{1}{\sqrt{2\pi}}\int_{-\infty}^x\exp\left(-\frac{\tau^2}{2}\right)\mathrm{d}\tau
\end{align*}
is the standard normal c.d.f., and $\mu_l$ and $\sigma_l$ are the log-normal parameters.

We plug \eqref{eq:approxt} and \eqref{eq:logncdf} into \eqref{eq:rexact} and, after rounding, arrive at
\begin{align}\label{eq:roundr}
\Bigg[\Phi\left(\frac{\log t - \mu_l}{\sigma_l}\right)(n + 1)\Bigg] \approx r,
\end{align}
where $[\cdot]$ denotes rounding to the nearest integer.

Maximum likelihood estimation (MLE) for the normal distribution yields log-normal estimators for $\mu_l$ and $\sigma_l$ as
\begin{align}\label{eq:mlestimates}
\left(\hat{\mu_l},\hat{\sigma_l}\right)=\left(\frac1c\sum_{i=1}^cq_i^{(l)},\sqrt{\frac1c\sum_{i=1}^c \left(q_i^{(l)} - \hat{\mu_l}\right)^2}\right)
\end{align}
by setting $q_i^{(l)}\coloneqq\log t_i^{(l)}$.

What remains to be done is finding an estimate for the total number of teams $n$ to estimate the scaling factor $(n+1)$ in \eqref{eq:roundr}. We assume that there are no ties, which is equivalent to stating that the elements in the training set $\bm{r}$ are unique. Thus, $\bm{r}$ is a length-$c$ sample, without replacement, of the discrete uniform distribution $\mathcal{U}[1,n]$.

Now recall that $\bm{r}$ corresponds to an ordered $B_c$ (an ordered proper $c$-subset of $\mathbb{N}_n$). Let $D$ denote a random variable that follows $\mathcal{U}[1,n]$. Estimating the parameter $n$ of $\mathcal{U}[1,n]$, with a sample drawn without replacement, is in the literature known as the \emph{German tank problem} \cite{ruggles1947}. A uniformly minimum-variance unbiased estimator (UMVUE) for this parameter is given in \cite{goodman1952} as

\begin{align}\label{eq:nestim}
\hat{n}=\left(1+\frac1c\right)r_{(c)} - 1,
\end{align}
where
\begin{align*}
r_{(c)}=\max_{i\in\mathbb{N}_c} r_i
\end{align*}
is the realization of the $c^\text{th}$ order statistic (maximum) of a length-$c$ sample of $D$.

We plug the pair $\left(\hat{\mu_l},\hat{\sigma_l}\right)$ of \eqref{eq:mlestimates} into \eqref{eq:roundr}. We plug \eqref{eq:nestim} into the $n$ of \eqref{eq:roundr}. This concludes the derivation.

\ifCLASSOPTIONcaptionsoff
  \newpage
\fi

\bibliographystyle{IEEEtran}
\bibliography{bibliography}

\begin{IEEEbiography}[{\includegraphics[width=1in,height=1.25in,clip,keepaspectratio]{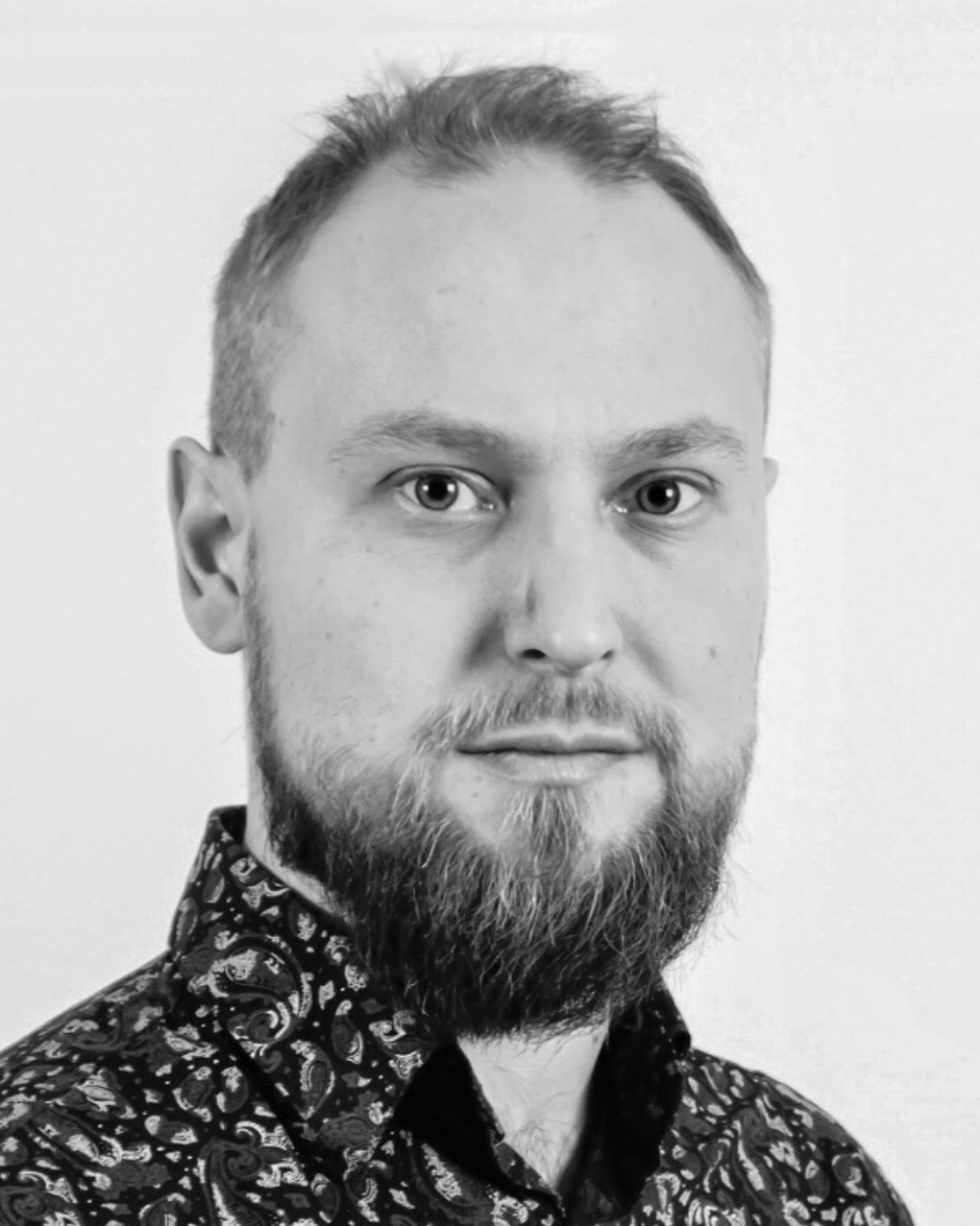}}]{Joonas~P\"a\"akk\"onen} received the MSc and PhD degrees in communications engineering from Aalto University, Finland, in 2012 and 2018, respectively. During his graduate studies, he worked as part of the Communications Theory Research Group under Prof. Olav Tirkkonen and as part of the Algebra, Number Theory and Applications Research Group under Prof. Camilla Hollanti. Later, he worked as the WOC Team Coach on the National Team of Orienteering Canada in 2019.

His research interests include wireless communications, distributed storage, coding theory, probability, and mathematical statistics. More recently, his research interests have expanded to numerical methods in sports science, as well as computational sports analytics, athlete training program planning, performance analysis and recovery analysis.

Dr. P\"a\"akk\"onen currently works as a researcher and lecturer at the Department of Informatics, School of Technology and Business Studies at Dalarna University, Borl\"ange, Sweden.
\end{IEEEbiography}

\end{document}